 \documentclass{emulateapj}

\slugcomment{To appear in ApJ December 20, 2007, v671n2 issue.}

\shorttitle{Search for Gamma-rays from SN 1987A}
\shortauthors{Enomoto et al.}

\begin{document}

\title{CANGAROO-III Search for Gamma Rays from SN 1987A
and the Surrounding Field}

\author{
R.~Enomoto\altaffilmark{1}
G.~V.~Bicknell\altaffilmark{2}
R.~W.~Clay\altaffilmark{3}
P.~G.~Edwards\altaffilmark{4}
S.~Gunji\altaffilmark{5}
S.~Hara\altaffilmark{6}
T.~Hattori\altaffilmark{7}
S.~Hayashi\altaffilmark{8}
Y.~Higashi\altaffilmark{9}
Y.~Hirai\altaffilmark{10}
K.~Inoue\altaffilmark{5}
S.~Kabuki\altaffilmark{9}
F.~Kajino\altaffilmark{8}
H.~Katagiri\altaffilmark{11}
A.~Kawachi\altaffilmark{7}
T.~Kifune\altaffilmark{1}
R.~Kiuchi\altaffilmark{1}
H.~Kubo\altaffilmark{9}
J.~Kushida\altaffilmark{7}
T.~Mizukami\altaffilmark{9}
R.~Mizuniwa\altaffilmark{7}
M.~Mori\altaffilmark{1}
H.~Muraishi\altaffilmark{12}
T.~Naito\altaffilmark{13}
T.~Nakamori\altaffilmark{9}
S.~Nakano\altaffilmark{9}
D.~Nishida\altaffilmark{9}
K.~Nishijima\altaffilmark{7}
M.~Ohishi\altaffilmark{1}
Y.~Sakamoto\altaffilmark{7}
A.~Seki\altaffilmark{7}
V.~Stamatescu\altaffilmark{3}
T.~Suzuki\altaffilmark{10}
D.~L.~Swaby\altaffilmark{3}
T.~Tanimori\altaffilmark{9}
G.~Thornton\altaffilmark{3}
F.~Tokanai\altaffilmark{5}
K.~Tsuchiya\altaffilmark{9}
S.~Watanabe\altaffilmark{9}
Y.~Yamada\altaffilmark{8}
E.~Yamazaki\altaffilmark{7}
S.~Yanagita\altaffilmark{10}
T.~Yoshida\altaffilmark{10}
T.~Yoshikoshi\altaffilmark{1}
Y.~Yukawa\altaffilmark{1}
}

\altaffiltext{1}{ Institute for Cosmic Ray Research, University of Tokyo, Kashiwa, Chiba 277-8582, Japan} 
\altaffiltext{2}{ Research School of Astronomy and Astrophysics, Australian National University, ACT 2611, Australia} 
\altaffiltext{3}{ School of Chemistry and Physics, University of Adelaide, SA 5005, Australia} 
\altaffiltext{4}{ Paul Wild Observatory, CSIRO Australia Telescope National Facility, CSIRO,
Narrabri, NSW 2390, Australia} 
\altaffiltext{5}{ Department of Physics, Yamagata University, Yamagata, Yamagata 990-8560, Japan} 
\altaffiltext{6}{ Ibaraki Prefectural University of Health Sciences, Ami, Ibaraki 300-0394, Japan} 
\altaffiltext{7}{ Department of Physics, Tokai University, Hiratsuka, Kanagawa 259-1292, Japan} 
\altaffiltext{8}{ Department of Physics, Konan University, Kobe, Hyogo 658-8501, Japan} 
\altaffiltext{9}{ Department of Physics, Kyoto University, Sakyo-ku, Kyoto 606-8502, Japan} 
\altaffiltext{10}{ Faculty of Science, Ibaraki University, Mito, Ibaraki 310-8512, Japan} 
\altaffiltext{11}{ Department of Physical Science, Hiroshima University, Higashi-Hiroshima, Hiroshima 739-8526, Japan} 
\altaffiltext{12}{ School of Allied Health Sciences, Kitasato University, Sagamihara, Kanagawa 228-8555, Japan} 
\altaffiltext{13}{ Faculty of Management Information, Yamanashi Gakuin University, Kofu, Yamanashi 400-8575, Japan} 

\begin{abstract}
Optical images of SN 1987A show a triple ring structure. 
The inner (dust) ring has recently increased in brightness and
in the number of hot spots suggesting that the supernova shock wave
has collided with the dense pre-existing circumstellar medium,
a scenario supported by radio and X-ray observations.
Such a shocked environment is widely expected to result in
the acceleration of charged particles, and the accompanying
emission of very high energy gamma-rays.
Here, we report the results of observations made in 2004 and 2006
which yield upper limits on the TeV gamma-ray flux, which
are compared with a theoretical prediction.
In addition, we set upper
limits on the TeV flux for four high energy objects which are 
located within the
same field of view of the observation: 
the super-bubble 30~Dor~C, the Crab-like pulsar PSR~B0540$-$69,
the X-ray binary LMC X-1, and the supernova remnant N157B.
\end{abstract}


\section{Introduction}

SN1987A was the closest observed supernova
in 380 years, and the evolution of the remnant of this supernova
has been studied in great detail over the past twenty years.
The radio intensity is growing and its rate of increase is
increasing, with the spectral index being observed to flatten
\citep{atnf}.
Imaging at radio wavelengths has enabled the expansion of the inner ring
to be traced, and is revealing increasing structure in the inner ring.
The X-ray fluxes observed with XMM and Chandra also
continue to rise almost exponentially \citep{xmm,chandra}. 
Around $\sim$4000 days after the supernova the X-ray
light curve increased dramatically, attributed to the
arrival of the supernova blast wave at the equatorial ring of
circumstellar material. There is some evidence that the X-ray flux
is mainly thermal, indicating the blast-wave interaction 
with dense matter \citep{chandra}.

\citet{berezhko} compared radio
and X-ray data with their calculations and concluded that there 
is a high (amplified) magnetic field inside the supernova remnant. 
The predicted TeV gamma-ray flux, resulting mostly from the
decay of neutral pions produced in interactions of the 
accelerated cosmic rays, is almost one order below the level
obtained in previous searches by 
Imaging Atmospheric Cherenkov Telescopes (IACTs) such as
CANGAROO-II in 2001 \citep{enomoto2003} 
and H.E.S.S.\ in 2003 \citep{rowell}. 
The wide field of view searched by H.E.S.S.\ did, however,
yield marginal excesses in a region north-east of SN 1987A and at LMC X-1.

CANGAROO-III is one of two stereoscopic 
IACTs located in the southern hemisphere. With a three-fold
coincidence, the sensitivity is significantly improved from the previous 
single telescope (CANGAROO-II). 
We also increased the observation period by a factor of two
in the 2006 observations compared to the 2004 and 2001 seasons.
We here report the results of these observations. 

The previous H.E.S.S.\ analysis considered not only SN 1987A but also
four other high energy objects within the same field of view \citep{rowell}.
We follow that idea and present upper limits on the integral
gamma-ray fluxes from 
the super-bubble 30~Dor~C, 
the Crab-like pulsar PSR B0540$-$69,
the X-ray binary LMC X-1, 
and the supernova remnant N157B.
Super-bubbles are expected to be efficient accelerators
of cosmic rays \citep{parizot} and
a shell feature near the star 30~Dor~C, within 5 arcmin of SN 1987A, shows
an indication of non-thermal hard X-ray emission \citep{bamba}.
The energy required to produce the bubble is estimated to be 
7$\times$10$^{51}$\,erg, which is
higher than generally expected for a single supernova remnant \citep{ueno}.
The other three sources belong to categories of
high-energy astrophysical objects that may
accelerate particles to TeV energies, but,
considering their large distance of 50\,kpc, they
are less likely to be detected
by current IACTs than their Galactic counterparts.
However, as they are well within the field of view
of the CANGAROO-III telescopes, and given the marginal H.E.S.S.\ excess
toward LMC X-1, it is straightforward to search
for any evidence of gamma-ray emission from them.

\section{CANGAROO-III Stereoscopic System}

The CANGAROO-III stereoscopic system consists of four imaging atmospheric
Cherenkov telescopes located near Woomera, South Australia (31$^\circ$S,
137$^\circ$E).
Each telescope has a 10\,m diameter segmented reflector 
consisting of 114 spherical mirrors 
each of 80\,cm diameter \citep{kawachi}, 
providing a total light collecting area of 57.3\,m$^2$.
The spherical segments are mounted on a parabolic
frame with a focal length of 8\,m.
The first telescope, T1, which was the CANGAROO-II telescope
\citep{enomoto_nature},
was not used for these observations due to its smaller field of view
and higher energy threshold.
The second, third, and fourth telescopes (T2, T3, and T4) were used for the
observations described here.
The camera systems for T2, T3, and T4 are identical and are described 
in \citet{kabuki}.
The telescopes are located at the 
east (T1), west (T2), south (T3) and north (T4)
corners of a diamond 
with sides of $\sim$100\,m \citep{enomoto_app}.

\section{Observations}

The observations were carried out 
in the periods from 2004 Nov 11 to 14 (MJD 53320--53323)
and from 2006 Dec 12 to 27 (MJD 54081--54096)
using ``wobble mode"
in which the pointing position of each telescope was
shifted in declination between $\pm$0.5 degree 
every 20 minutes \citep{wobble}
from the target:
(RA, dec [J2000]) = (83.866139$^\circ$, $-$69.269577$^\circ$).  
We took no OFF source runs.
The most sensitive region of the field of view
is within one degree of the average pointing position.
LMC~X-1 is located 0.6 degrees south-east from SN~1987A.
Considering our angular resolution of 0.24 degree, there is
some overlap between  
SN 1987A, 30~Dor~C, PSR B0540$-$69, and N157B.

In the 2004 observation,
data with GPS time stamps were recorded for T2, T3 and T4 individually when
more than four photomultiplier (PMT) signals 
exceeded 7.6 photoelectrons (p.e.). 
In the offline analysis stage we combined the data when
the three telescope's GPS times coincided.
In the 2006 observation, a hardware coincidence was used to select 
any two triggered telescopes \citep{nishijima}.
The images in all three telescopes were required to have clusters
of at least five adjacent pixels exceeding a 5\,p.e.\ threshold
(offline three-fold coincidence).
The event rate was reduced to 5$\sim$8\,Hz by this criterion
depending on the elevation angle and year.
Based on time dependence of these rates we can remove data
taken in cloudy conditions. 
The effective observation times 
for 2004 and 2006 were
632 and 1316 min, respectively.
The corresponding mean zenith angles were 42.1$^\circ$ 
and 40.0$^\circ$.

The light collecting efficiencies, including the reflectivity
of the segmented mirrors, the light guides, and the quantum efficiencies
of the photomultiplier tubes were monitored by a muon-ring analysis
\citep{enomoto_vela} with the individual trigger data in the
same periods. 
The light yield per unit arc-length is approximately proportional
to the light collecting efficiencies.
Deterioration is mostly due to dirt and dust settling on the
mirrors.

\section{Analysis}

Here, we briefly describe the analysis procedures, which 
are identical with those described in \citet{cena}.
More details can be found in \citet{enomoto_vela} and
\citet{enomoto_0852}.

First, the image moments \citep{hillas} 
were calculated for the three telescopes.
The incident gamma-ray direction was determined by minimizing
the sum of squared widths ($\chi^2_0$: weighted by the photon yield) 
of the three images seen from the assumed position (fitting parameter)
with a constraint on the distances from the intersection point to each
image center.

In order to derive the gamma-ray likeliness,
we used  
the Fisher Discriminant (hereafter $FD$) \citep{fisher,enomoto_vela}.
Input parameters were
$$\vec{P}=(W2,W3,W4,L2,L3,L4),$$
where $W2,W3,W4,L2,L3,L4$ are energy-corrected $widths$ and $lengths$ for the
T2, T3, and T4
and assume that a linear combination of
$$FD=\vec{\alpha}\cdot\vec{P},$$
provides the 
best separation between signal and background, then the set of
linear coefficients ($\vec{\alpha}$) should be uniquely determined as
$$\vec{\alpha}=\frac{\vec{\mu}_{sig}-\vec{\mu}_{BG}}{E_{sig}+E_{BG}},$$
where $\vec{\mu}$ is a vector of the mean value of $\vec{P}$ for each
sample and $E$ is their correlation matrix.
We previously used 
it in Vela Pulsar analysis \citep{enomoto_vela}
to separate ``sharp'' (gamma-ray--like) images from ``smeared'' (background) 
ones.
The values of
$\vec{\mu}_{sig}$, $\vec{\mu}_{BG}$,
$E_{sig}$, and $E_{BG}$ can be calculated from the
Monte-Carlo and observational data (OFF-source runs), respectively.

We rejected events with any hits in the outermost layer of a camera (``edge
cut"). These rejected events suffer from finite deformations, especially in
the $length$ distribution, which results in deformations of the $FD$.

We then derived $FD$ distributions position by position.
Comparing those in the signal region and control background
region, we can determine the gamma-ray--like events.
Here, we use Monte-Carlo simulations to determine the $FD$ 
distribution of gamma-ray signals.
Note that in the gamma-ray simulations we used a spectrum 
proportional to $E^{-2.1}$.
The fit of the $FD$ distribution of source position
with the above emulated
signal and control background functions were carried out, 
to derive the number of gamma-ray--like events.
This is a one-parameter fitting with the constraint that
sum of signal and background events corresponds to the total number of events.
These coefficients can be derived exactly analytically.

\section{Results}

\begin{figure*}[htbp]
\includegraphics[width=300pt]{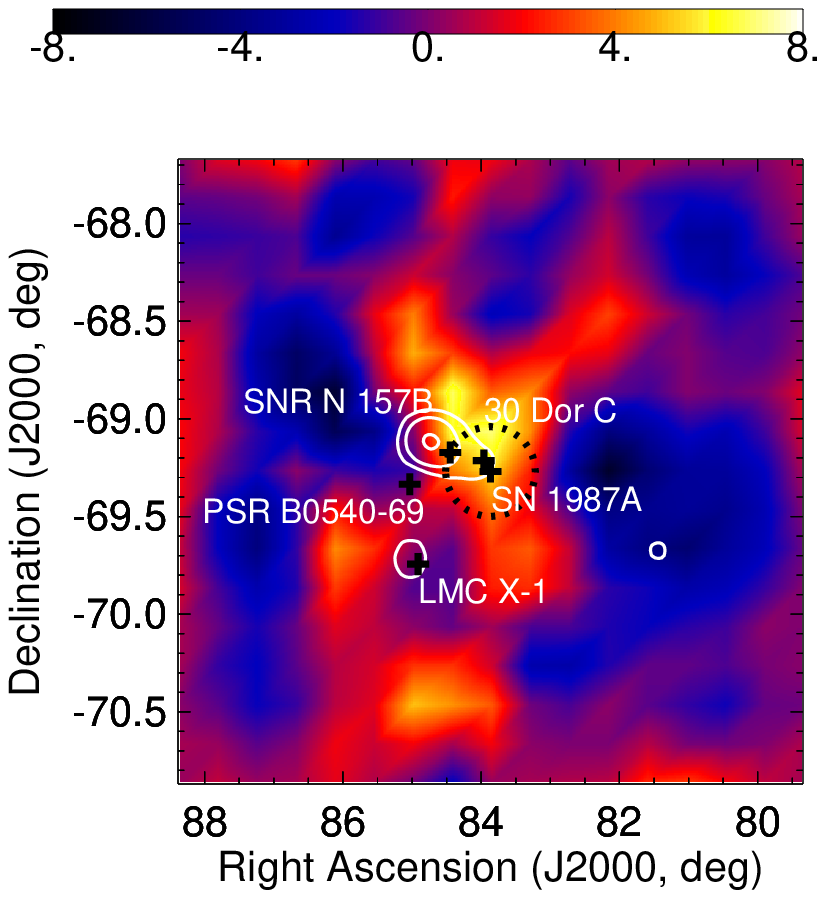}
\hskip -3cm\includegraphics[width=300pt]{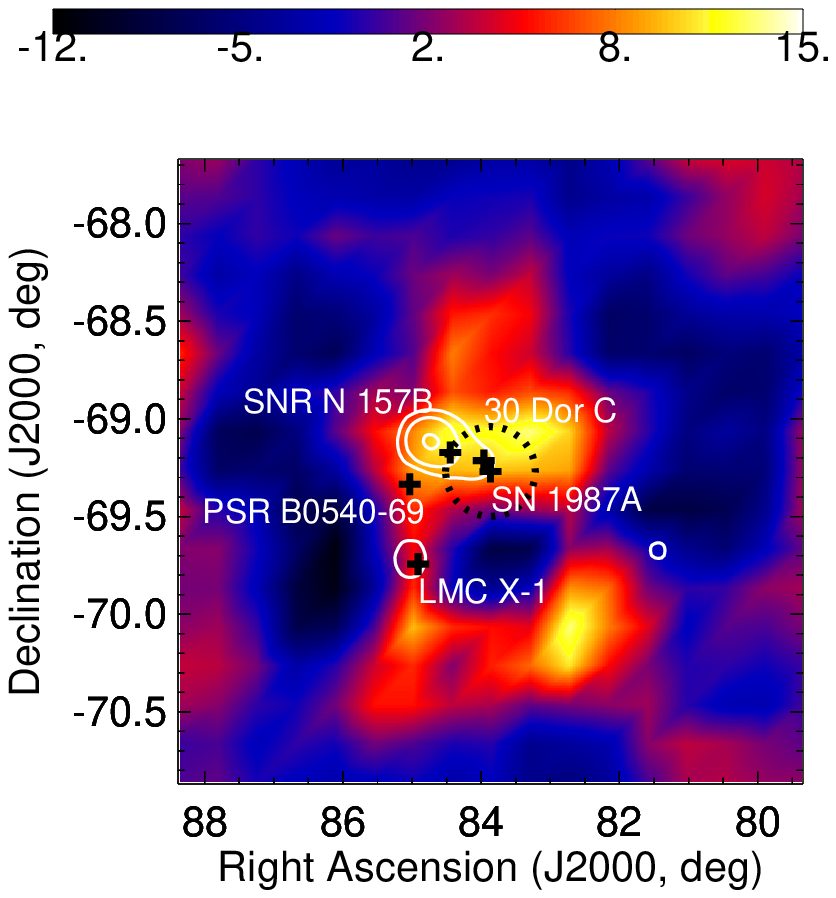}
\vskip -1cm
\caption{Excess count maps. The left panel is obtained from the 2004 data and
the right from the 2006 data. The average pointing center is SN 1987A itself.
The positions of four other high energy objects are also 
indicated by the black
crosses with labels. 
The PSF is shown by the black dashed circle.
The white contours are derived from 
4.85\,GHz PMN survey radio data \citep{pmns}
obtained from Skyview (NASA) \citep{skyview}.}
\label{fig1}
\end{figure*}

The threshold of this analysis is considered to be $\sim$1\,TeV
which is higher than the typical CANGAROO-III threshold due to
the larger zenith angle of the observations, of around 40 degrees.
In order to derive morphology, we segmented the field of view into
0.2\,$\times$\,0.2 degree$^2$ square bins. The $FD$ distributions
for corresponding bins are made and fitted. The control-background
region is defined as 
the second closest layer,
which is more than 0.3
degree from the center of target region,
i.e., larger than the 0.24 degree point spread function (PSF). 
The statistics of the
control-background is, therefore, sixteen times larger than that
of signal bin.
The results are shown in Figs.~\ref{fig1}.
The left panel is obtained from the 2004 data and the right from 2006.
The average telescope pointing position is indicated by the
cross centered and labeled as ``SN 1987A", with the other 
four high energy objects within
this field of view indicated by crosses.
Our sensitivity is considered to be limited up to
one degree in radius from the center. The 
PSF is considered to be a 0.24 degree radius circle.
The significance distributions (excess divided by the statistical error)
are approximately normal (Gaussian) distributions.
The best fit Gaussians have mean values 
of $-$0.31$\pm$0.08 (2004) and $-$0.18$\pm$0.09 (2004),
and standard deviations of 1.21$\pm$0.06 (2004) and 1.31$\pm$0.06 (2006),
within systematic uncertainties.

\begin{figure}[htbp]
\plotone{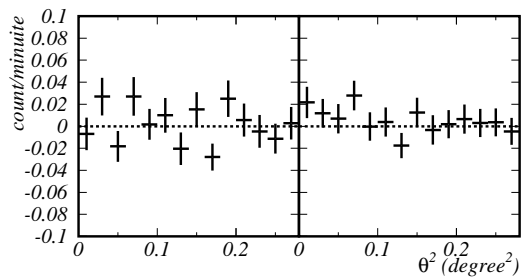}
\caption{Distributions of $\theta^2$ [degree$^2$]. The left panel
is obtained from the 2004 data and the right from the 2006 data.
The vertical scale is normalized to counts/minute per bin
[0.2 degree$^2$]. The
background-subtracted signals (obtained by the fitting
procedure described in the text) are plotted.}
\label{fig2}
\end{figure}

At first we investigate the region of SN 1987A. Note that
this region at some energies is contaminated by 30~Dor~C, 
a highly energetic source.
To indicate the excess distribution from its center, we show
so-called $\theta^2$ distributions in Figs. \ref{fig2}.
Here, the control-background sample was selected in the region
$\theta^2$=(0.1--0.2) [degree$^2$].
The background-subtracted signals are shown.
Both 2004 and 2006 data have statistically
insignificant excesses near $\theta^2=0$.
The $\chi^2/$DOFs (degree of freedom) for null assumptions are 18.6/15
and 13.2/15 for 2004 and 2006, respectively.
These statistically insignificant excesses 
appear as the small excess count peaks around the centers of 
the plots in Fig.~\ref{fig1}.
Note that our PSF corresponds to $\theta^2 <$0.06 degree$^2$.
We, therefore, proceed assuming that there is no signal present.
In order to show the consistency between 2004 and 2006, the
vertical unit is unified to count rate.

\begin{figure}[htbp]
\plotone{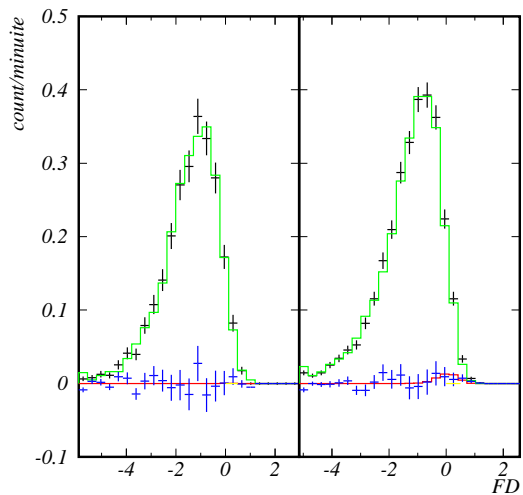}
\caption{$FD$ distributions. The left panel is obtained from
the 2004 data and the right the 2006 data. The vertical
scale is counts/minute per bin [arbitrary unit]. The black
crosses are obtained from the on-source region (centered
around SN 1987A within the PSF), the green histograms the best fit
of the control backgrounds, the blue crosses the
background-subtracted signals, and the red histograms
the best fit signal distributions.}
\label{fig3}
\end{figure}

In order to determine whether or not there is a
gamma-ray excess around SN 1987A, 
we made the
$FD$ distributions within the PSF. The control-background was made
from the region $\theta^2$=(0.1--0.2). The fitting results are
shown in Fig. \ref{fig3}.
The background level differs between the two years.
This is due to sky conditions, mirror reflectivity, time-dependence of
the blur spot size (which improves after maintenance periods), etc.
These are considered to be the major systematics for estimating the
observational gamma-ray flux in this work. 
For the light correction efficiency, we may have 10\%
uncertainty at maximum.

Again there is no statistically significant excess.
We carried out a similar analysis, year by year, energy
threshold by energy threshold.
The obtained 2$\sigma$ upper limits (ULs) are listed in Table~\ref{table1}.
\begin{table}[htbp]
\caption{The 2$\sigma$ upper limits to the integral fluxes 
from SN1987A
at two epochs with three different energy thresholds.}
\label{table1}
\begin{tabular}{cccc}
\hline\hline
Year & Excess Upper Limit & Energy Threshold & Flux Upper Limit \\
& Events & GeV & ${\rm cm}^{~2}{\rm s}^{-1}$\\
\hline
2004 & \,35\,~  &  1200  &   1.7 $\times$ 10$^{-11}$ \\ 
2004 & \,21\,~  &  2870  &   5.7 $\times$ 10$^{-13}$ \\ 
2004 & \,~4.6   &  7180  &   1.0 $\times$ 10$^{-13}$ \\ 
2006 & 116\,~   &  1080  &   2.7 $\times$ 10$^{-12}$ \\
2006 & \,19\,~  &  2550  &   2.5 $\times$ 10$^{-13}$ \\
2006 & \,~5.3   &  6350  &   6.1 $\times$ 10$^{-14}$ \\
\hline\hline
\end{tabular}
\end{table}
Here we used a $E^{-2.1}$ spectrum for the gamma-ray
simulation.
The ULs range from 6--17\% crab. The worse upper limits in the lower
energy range originated from the statistically insignificant
excess around the center of the field of view. At higher energies, 
we do not see any excess.

The four other high energy sources in the surrounding region, 
30~Dor~C, PSR B0540$-$69, LMC X-1, and N157B, can be
analyzed in the same way. They are all within a one degree circle
from the average pointing position (the sensitive region).
The summary is listed in Table~\ref{table2}.
\begin{table}[htbp]
\caption{The 2$\sigma$ upper limits to the integral fluxes at energy 
greater than 1120~GeV for five high energy objects.
The 2004 and 2006 data were averaged.}
\label{table2}
\begin{tabular}{cccc}
\hline\hline
Target & Excess Upper Limit & Flux Upper Limit \\
& Events & ${\rm cm}^{-2}{\rm s}^{-1}$\\
\hline
SN 1987A       & 126 & 1.8 $\times$ 10$^{-12}$ \\ 
30 Dor C       & 165 & 2.5 $\times$ 10$^{-12}$ \\ 
PSR B0540$-$69 & ~66 & 1.1 $\times$ 10$^{-12}$ \\ 
LMC X-1        & ~74 & 1.1 $\times$ 10$^{-12}$ \\ 
N157B          & 167 & 2.4 $\times$ 10$^{-12}$ \\ 
\hline\hline
\end{tabular}
\end{table}
These are two year averages.
Note that, considering our PSF of 0.24 degree, SN 1987A and
30~Dor~C are highly confused, with other pairs being less confused.
Also some of the control-background regions are confused.
The acceptance is a slowly and smoothly decreasing function and is reduced
to 65\% at a distance of one degree from the center.

\section{Discussion}

Our morphology indicates a negligibly 
small excess near the center of field of
view both in the 2004 and 2006 data. This is statistically $<$2$\sigma$
each year.
We note that a similar feature can also be seen in the H.E.S.S.\ data shown 
in Figs.~2 and 3 of \citet{rowell}. A deep survey,
i.e., longer observation, around this region is awaited.
The region around LMC X-1 is consistent with zero in our data.

\begin{figure}
\plotone{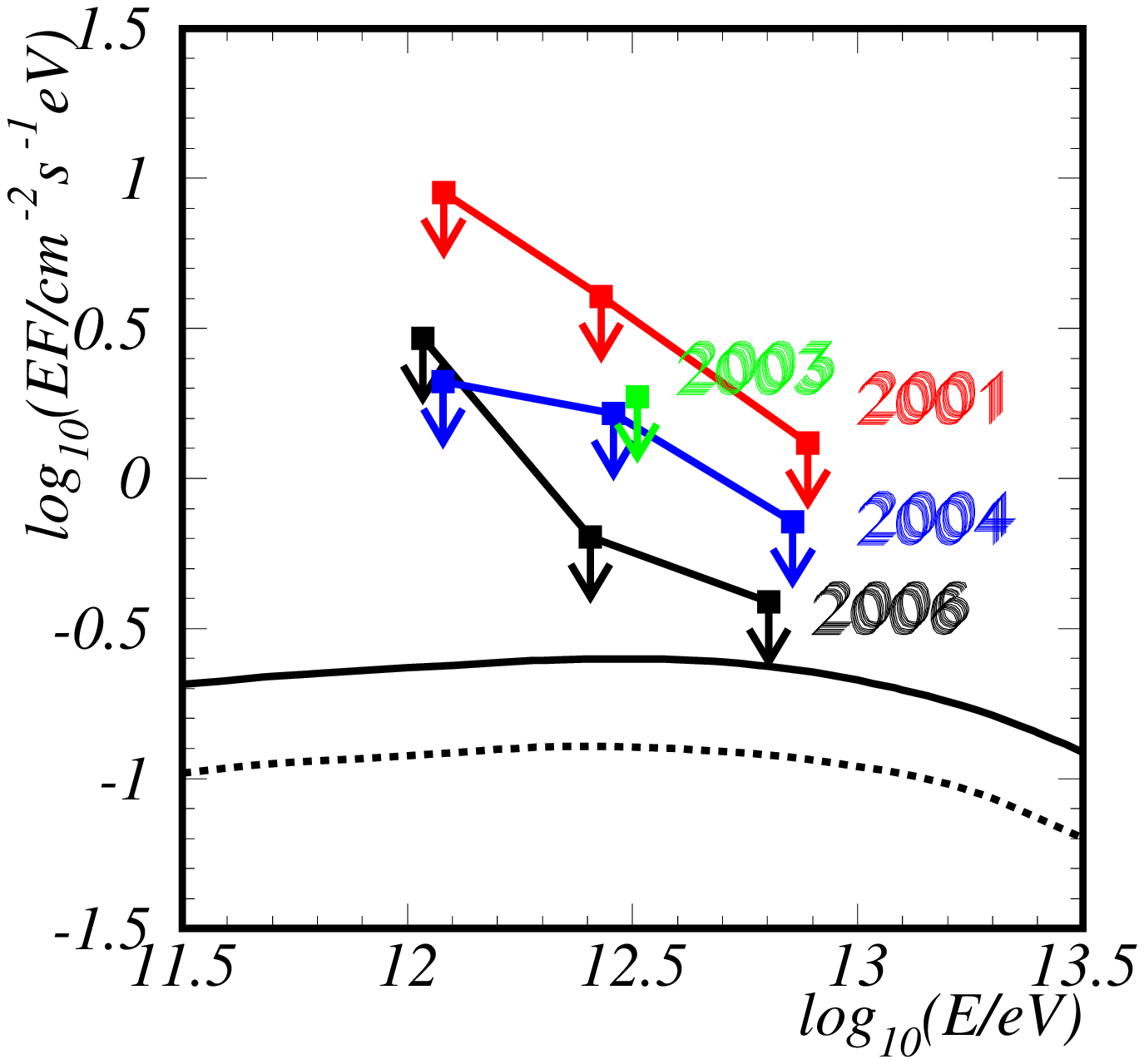}
\caption{Spectral energy distribution.
The blue points and line are obtained from the 2004 data
and the black from the 2006 data. The red points are the CANGAROO-II
upper limits from 2001 data \citep{enomoto2003}. The green points are
the H.E.S.S.\ data from 2003 \citep{rowell}.
The lines are obtained from 
Fig.\ 4 of \cite{berezhko}. The solid and dashed curves are
the predictions of the gamma-ray flux 8249 and 7300
days after the supernova, respectively.}
\label{fig4}
\end{figure}

The upper limits of SN 1987A in Table~\ref{table1} 
are plotted in Fig.~ \ref{fig4}.
The results presented here (in blue and black) are compared
with CANGAROO-II (red)
\citep{enomoto2003} and H.E.S.S.\ (green)
\citep{rowell} results. The upper limits in this work are 
improved by factors of up to 10 on the 2001 result.
That is consistent with the 
longer observation time and better signal-to-noise ratio resulting from 
the three-fold coincidence.
However, the theoretical prediction even 7300 days after the supernova 
(the dashed curve) \citep{berezhko} is still a factor of three
below the results of the observations.
Note that  
the predictions \citep{berezhko} were made
in the frame of spherical symmetric approach. The existence of
strongly asymmetric interstellar matter structure,
which is a dense inner ring, makes
their prediction of gamma-ray flux uncertain. According to their rough
estimation, this uncertainty is not very large (about a factor of 2) due
to the stronger supernova shock deceleration in the denser medium.
The solid curve is their prediction for 8249 days after the supernova, still
lower than the upper limits obtained in this work.
In order to constrain this model, we need further improvements 
on the hardware and/or analysis methods.
A deeper observation by H.E.S.S., such as T$>$30 h, around $\sim$2010
might give conclusive results.
Also proposed future large-scale IACT arrays such as CTA \citep{cta} 
would be very promising.

The limit obtained for 30~Dor~C in Table~\ref{table2} is about
15\% crab level at $\sim$1\,TeV.
Considering the distance of 50\,kpc to LMC, this upper limit
corresponds to a huge cosmic-ray density comparing with that of 
the Crab nebula (distance of 2\,kpc).
Even if there is supernova complex there, the expected
TeV gamma-ray flux is still below our sensitivity limit.
Increasing the cosmic-ray target density such 
as either inter-stellar-matter
or photon density within plausible ranges, the expected flux  
does not exceed this high upper limit.
According to the standard estimation on the hadronic gamma-ray production
such as \citet{drury,naito} with an assumption of the target density of
1p/cm$^{-3}$, a total kinetic explosion energy of $7\times10^{51}$\,ergs
gives a 1 milli-crab flux of gamma-rays at the solar system.
Again future larger-scale projects are required to probe such flux levels.

For the other three high energy objects, we set upper limits
far below those of previous observations.
We, however, could not find 
any physically important constraints on
their activity at high energies again
due to the distance of 50\,kpc.

\section{Conclusion}

We have observed SN 1987A at two epochs: MJD 53320--53323
(2004 Nov) and MJD 54081--54096 (2006 Dec), 
approximately 6500 and 7200 days after the supernova, respectively.
The effective observation times are 10.5 and 20\,h, respectively. 
No statistically significant gamma-ray signals were obtained
and we set 2$\sigma$ upper limits on the integral fluxes of 6--17\%
crab at $\sim$1\,TeV.
These are slightly lower than the previous CANGAROO-II and H.E.S.S.\
upper limits.
Although theory predicts a 
factor of $\geq$3 lower flux 7300 days after supernova,
it also predicts that the future flux level might exceed the  
sensitivity limit of the existing and future arrays.
Continued monitoring of this object at TeV energies 
is still therefore meaningful in constraining theoretical models.
It is also important to improve the sensitivity
of the observation methods, as proposed by the CTA project \citep{cta}.

In addition, we set 2$\sigma$ upper limits for 
four more high energy objects inside our
field of view: 
30~Dor~C (super bubble), PSR B0540$-$69 (Crab-like pulsar),
LMC X-1 (X-ray binary), and N157B (supernova remnant).

\acknowledgments

We thank L.T.\ Ksenofontov for discussion on the estimated gamma-ray
flux from SN 1987A.
This work was supported by a Grant-in-Aid for Scientific Research by
the Japan Ministry of Education, Culture, Sports, Science and Technology, 
the Australian Research Council, JSPS Research Fellowships,
and the Inter-University Research Program 
of the Institute for Cosmic Ray Research.
We thank the Defense Support Center Woomera and BAE Systems.

\end{document}